\documentclass[journal]{IEEEtran}

\usepackage{graphicx}
\usepackage{lineno}
\usepackage{stfloats}
\usepackage{textcomp}

\ifCLASSINFOpdf

\else

\fi

\begin{document}

\title{Silicon single-photon detector achieving over 84\% photon detection efficiency with flexible operation modes}

\author{Dong~An,
         Chao~Yu, 
	Ming-Yang~Zheng,
         Anran~Guo, 
	Junsong~Wang,
	Ruizhi~Li,
	Huaping~Ma,
	Xiu-Ping~Xie,
	Xiao-Hui~Bao,
	Qiang~Zhang,
	Jun~Zhang, and 
	Jian-Wei~Pan
\thanks{Dong An, Chao Yu, and Junsong Wang are with Hefei National Research Center for Physical Sciences at the Microscale and School of Physical Sciences, University of Science and Technology of China, Hefei 230026, China, and are also with CAS Center for Excellence in Quantum Information and Quantum Physics, University of Science and Technology of China, Hefei 230026, China (e-mail: yuch@ustc.edu.cn).}
\thanks{Ming-Yang Zheng, and Xiu-Ping Xie are with Jinan Institute of Quantum Technology and CAS Center for Excellence in Quantum Information and Quantum Physics, University of Science and Technology of China, Jinan 250101, China, and are also with Hefei National Laboratory, University of Science and Technology of China, Hefei 230088, China.}
\thanks{Anran Guo, Ruizhi Li and Huaping Ma are with Department of Solid State Image Sensor, CETC No.44 Research Institute, China, and are also with CETC Chips Technology Group CO., LTD, Chongqing 401332, China.}
\thanks{Qiang Zhang is with Hefei National Research Center for Physical Sciences at the Microscale and School of Physical Sciences, University of Science and Technology of China, Hefei 230026, China, and with Jinan Institute of Quantum Technology and CAS Center for Excellence in Quantum Information and Quantum Physics, University of Science and Technology of China, Jinan 250101, China, and is also with Hefei National Laboratory, University of Science and Technology of China, Hefei 230088, China.}
\thanks{Xiao-Hui Bao, Jun Zhang and Jian-Wei Pan are with Hefei National Research Center for Physical Sciences at the Microscale and School of Physical Sciences, University of Science and Technology of China, Hefei 230026, China, and with CAS Center for Excellence in Quantum Information and Quantum Physics, University of Science and Technology of China, Hefei 230026, China, and also with Hefei National Laboratory, University of Science and Technology of China, Hefei 230088, China (e-mail: zhangjun@ustc.edu.cn).}}

\maketitle

\begin{abstract}

Silicon single-photon detectors (Si SPDs) play a crucial role in detecting single photons in the visible spectrum. For various applications, photon detection efficiency (PDE) is the most critical characteristic for effectively collecting photons. Here, we present a Si SPD with a remarkable PDE of up to 84.4\% at 785 nm, supporting multiple operation modes. We design and fabricate a thick-junction Si single-photon avalanche diode (SPAD) that enhances the avalanche probability through a backside-illumination structure, while minimizing noise through the design of a doping-compensated avalanche region. To maximize PDE, we implement a readout circuit with a 50 V quenching voltage, enabling operation in free-running, gating, or hybrid modes. The SPAD, along with its readout circuits and affiliated circuits, is integrated into a compact SPD module. In free-running mode, the module achieves a maximum PDE of 84.4\%, with a dark count rate of 260 cps, and an afterpulse probability of 2.9\% at 268 K. This work provides a practical solution for applications requiring ultra-high-efficiency Si SPD with multiple operation modes. 

\end{abstract}


\section{Introduction}

\IEEEPARstart{S}{ilicon} single-photon detectors (Si SPDs)~\cite{HPB93} are widely used for detecting single photons in the visible spectrum. Owing to their compact size, low cost, and ease of operation, Si SPDs have play a pivotal role in diverse applications such as quantum photonics~\cite{FGA21} and single-photon imaging~\cite{GNR15}. 
Key performance metrics of Si SPDs include photon detection efficiency (PDE), dark count rate (DCR), afterpulse probability, timing jitter, and maximum count rate. In most applications, high PDE is crucial for effectively collecting photons. For instance, in a ten-photon entanglement experiment, a 5\% relative improvement in PDE can lead to a 63\% increase in coincidence count rate~\cite{JZC12,XLW16}. As a result, improving PDE become a central objective in the development of Si SPDs.

\begin{figure*}[t]
\centering
\includegraphics[width=16 cm]{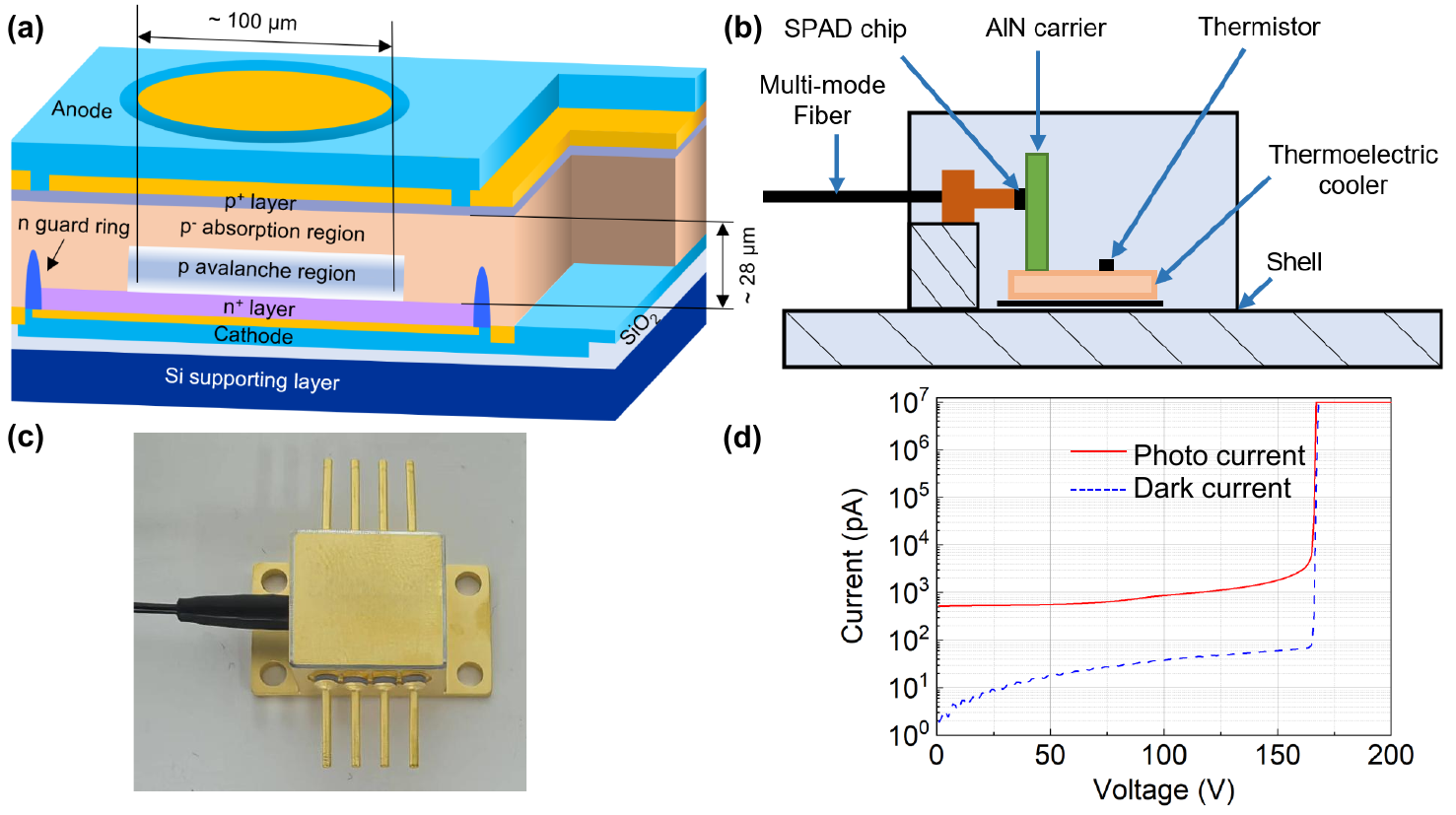}
\caption{(a) Schematic diagram of the Si SPAD semiconductor structure. (b) Schematic of Si SPAD butterfly package. (c) Photograph of the Si SPAD device. (d) Measured I-V curve of the Si SPAD.}
\label{fig1}
\end{figure*}

In Si SPDs, PDE is primarily determined by two components: the single-photon avalanche diode (SPAD) and the associated readout circuit~\cite{SMA96,SMA04}. Currently, three main types of Si SPADs have been reported: complementary metal-oxide-semiconductor (CMOS) SPADs~\cite{DSF13}, thin-junction SPADs~\cite{AFI10}, and thick-junction SPADs~\cite{SJY99}. CMOS and thin-junction SPADs offer advantages in integration and low timing jitter, but their thin absorption layers result in relatively low PDE. Although certain nanostructure designs can enhance PDE~\cite{JMZ15}, these devices still underperform in the near-infrared region, reaching only around 32\% PDE at 850 nm~\cite{KXY17}. In contrast, thick-junction SPADs, featuring thick absorption layers of tens of micrometers, exhibit superior photon absorption and higher PDE. For instance, commercially available thick-junction Si SPDs~\cite{SPCM} demonstrate excellent PDE of $\sim$70\% at 780 nm. Nevertheless, fabricating a thick-junction Si SPAD with a PDE of over 80\% remains a highly challenging task.

For a given Si SPAD, PDE is primarily influenced by the excess bias voltage $V_{ex}$ ($V_{ex}=V_{b}-V_{br}$), where $V_{b}$ is the bias voltage and the $V_{br}$ is the breakdown voltage of SPAD. In thick-junction Si SPADs, $V_{ex}$ typically exceeds 30 V to ensure a high avalanche probability. Further increasing the excess bias voltage can push the PDE to its limit. For instance, by applying a sine-wave gating circuit with a 50 V peak-to-peak amplitude~\cite{NWL17} and a monolithic integrated circuit with a 68 V quenching voltage~\cite{YKX20} to SPADs disassembled from commercial products, PDE was enhanced from 68.6\% to 73.1\% and from 69.5\% to 75.1\% at 785 nm, respectively. 

In addition to increasing the quenching voltage to achieve the maximum efficiency, flexible operation modes are also a crucial requirement for practical applications. Based on the different quenching approaches, the operational modes of the Si SPD can be categorized into free-running mode and gating mode. In free-running mode~\cite{GIA16,MBP17,GII17,GMI18,GAM18}, the SPD can detect asynchronously arriving photons, with the quenching signal being triggered by the avalanche event after a short, fixed transmission delay. In gating mode~\cite{OZJ10,SNK14}, the SPD detects photons only within the gate-on period, which offers advantages such as reduced noise and high count rates. The avalanche signal is passively quenched at the end of the gate. In specific scenarios, a hybrid approach combining both modes, which means detecting signal photons in free-running mode during the gate-on period, is necessary to detect asynchronously incoming signal photons while suppress noise.

In this paper, we present a Si SPD with a peak PDE of 84.4\% at 785 nm and support for multiple operation modes. We design and fabricate a thick-junction Si SPAD with optimized semiconductor structure. A readout circuit  with an amplitude of 50 V is developed to maximize the PDE, enabling operation in free-running, gating, and hybrid modes. The SPAD, readout circuits, and affliated circuits are integrated into a compact SPD module with dimensions of 9 cm $\times$ 10 cm $\times$ 3 cm. Calibration in free-running mode shows the module achieves 84.4\% PDE at 785 nm, with a DCR of 260 cps, afterpulse probability of 2.9\%, and timing jitter of 360 ps at 268 K. 

\section{Silicon SPAD}

The schematic cross-section of the SPAD chip is shown in Fig.~\ref{fig1}(a), which consists of a $p^{+}$ contact layer, a $p^{-}$ absorption region, a p-type avalanche region, an $n^{+}$ contact layer, and an n-type guard ring. Following the method in\cite{AHR24}, we design a doping-compensated avalanche region by successive boron and phosphorus diffusion. This process compensates doping in the shallow region and concentrates the electric field in the deeper region, where fewer crystal defects are present, thus reducing noise. Additionally, a backside-illumination structure is employed to enhance the SPAD's PDE. Since the intensity of incident photons decays exponentially in silicon, this design ensures that most of the incident photons are absorbed in the absorption layer, with the generated electrons injected into the avalanche region. In contrast, in a front-illumination structure, photons absorbed in the surface layer inject holes into the avalanche region. In silicon, the ionization rate of electrons is higher than that of holes, meaning the avalanche probability for electron injection is greater than that for hole injection. Therefore, the backside-illumination design enables higher detection efficiency.

The fabrication begins with a 6-inch $p^{-}$/$p^{+}$ epitaxial silicon wafer. An n-type guard ring is first formed around the p-type avalanche region via phosphorus implantation to suppress edge effects and prevent edge prebreakdown. The avalanche region is then created through a two-step implantation diffusion process. Initially, low-energy boron implantation minimizes lattice damage, followed by high-temperature diffusion at 1200$^{\circ}$C to drive boron into the epitaxial layer. Phosphorus implantation and diffusion follow, further redistributing boron to establish the desired doping profile. After the avalanche region is formed, an $n^{+}$ contact layer is diffused. Subsequently, a contact hole is etched, and the cathode is sputtered. After completing the front-side process, the silicon wafer is flipped and bonded to the supporting silicon wafer. Then, the substrate is completely removed. After that, a $p^{+}$ layer is injected onto the backside, followed by the growth of an anti-reflection coating, the opening of contact holes, and the growth of the backside metal electrode. Finally, deep silicon etching is used to create a through-hole in the silicon wafer, exposing the front electrode for wire bonding. The completed SPAD has an depletion layer thickness of $\sim$ 28 µm and an active area diameter of $\sim$100 µm.

A butterfly package is employed to assemble the device, as schematically illustrated in Fig.~\ref{fig1}(b). The Si SPAD chip is bonded to a metallized aluminum nitride (AlN) carrier substrate, which is vertically mounted on a thermoelectric cooler (TEC). The TEC regulates control the device's operation temperature, with a thermistor placed on the top to provide temperature feedback. For efficient photon coupling, a multi-mode fiber is precisely aligned with the center of the SPAD chip. The final device package is shown in Fig.~\ref{fig1}(c). The packaged SPAD is characterized at 273 K, and its current-voltage (I-V) characteristics are presented in Fig.~\ref{fig1}(d). The breakdown voltage ($V_{br}$) is measured to be 170 V, and the dark current is approximately 70 pA when the bias voltage is 1 V below $V_{br}$.

\section{Readout circuit}

\begin{figure*}[htbp]
\centerline{\includegraphics[width=18 cm]{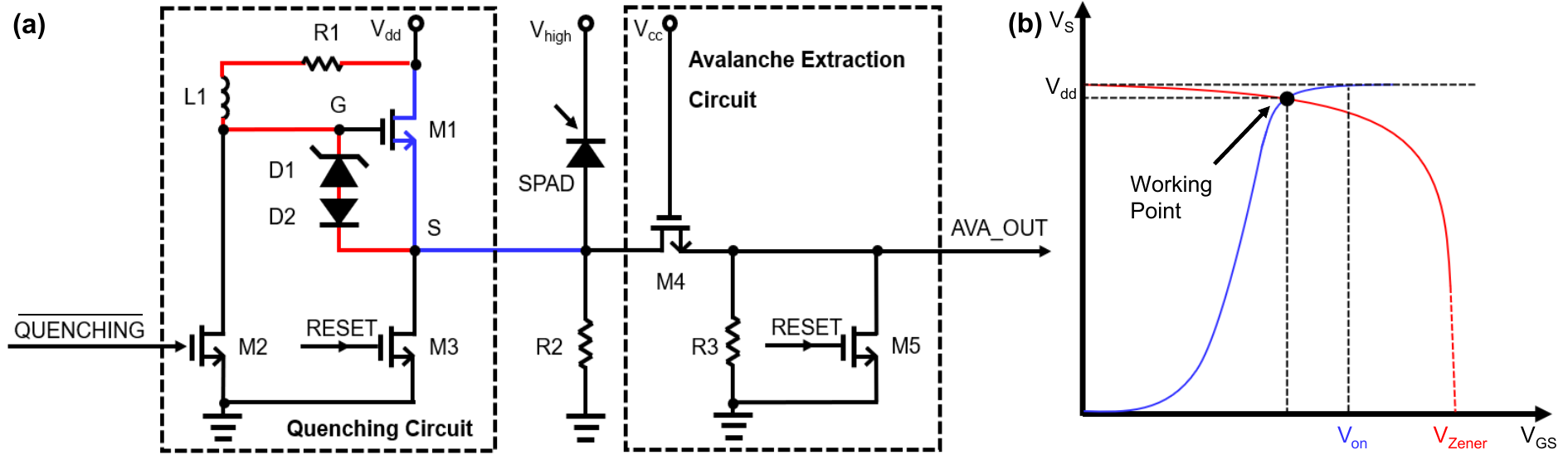}}
\caption{(a) Schematic diagram of the readout circuit. (b) The idle state working point estimation of the quenching circuit.}
\label{fig2}
\end{figure*}

Fig.~\ref{fig2} (a) shows the schematic of the readout circuit, which comprises a quenching circuit and an avalanche extraction circuit. The SPAD cathode is biased with a positive voltage, while the anode is connected to ground (GND) through a 20 k$\Omega$ resistor (R2). The output of the quenching circuit and the input of the avalanche extraction circuit are both connected to the SPAD anode. A field-programmable gate array (FPGA) provides the QUENCHING and RESET control signals, enabling flexible operation modes.

The quenching circuit can switch between a high-voltage mode and a high-resistance mode. In the high-voltage mode, the SPAD's reverse bias drops below its breakdown threshold, placing it in an idle state. In contrast, the high-resistance mode pulls the SPAD cathode to 0 V, setting the device in its armed state. The quenching circuit includes three n-channel double-diffused metal-oxide-semiconductor (DMOS) transistors (M1–M3), a high-voltage power supply ($V_{dd}$), a Zener diode (D1), a standard diode (D2), a 1 k$\Omega$ current-limiting resistor (R1), and an inductor (L1). The circuit operation is analyzed under four distinct conditions: armed state, idle state, transition from armed to idle, and transition from idle to armed.

In the armed state, the QUENCHING signal is set to a logic high (HIGH), while the RESET signal remains a logic low (LOW). This turns on M2 and turns off M3, pulling the voltage at point G ($V_{G}$) to GND, which turns M1 off. The reverse-connected diode D2 blocks avalanche current from flowing from the SPAD anode through M2 to GND. Consequently, the quenching circuit presents high impedance, and the SPAD anode is pulled to GND through resistors R2 and R3.

In the idle state, the QUENCHING signal is set LOW, and the RESET signal remains LOW. M2 turns off, allowing current from $V_{dd}$ to flow through D1 and D2. This increases the gate-to-source voltage ($V_{GS}$) of M1, partially turning it on. Fig.\ref{fig2} (b) provides a working point analysis for M1. In the branch circuit $V_{dd}$-R1-L1-D1-D2, the voltage at point S is given by $V_{S} = V_{dd} - i_{1} * R1 - V_{GS}$, where $i_{1}$ is the current through the branch, which depends on $V_{GS}$ and is determined by the volt-ampere characteristics of D1. At low voltages, $i_{1}$ increases slowly with $V_{GS}$. As $V_{GS}$ approaches the Zener voltage ($V_{Zener}$) of D1, $i_{1}$ increases sharply, causing a steep drop in $V_{S}$. The red curve in Fig.\ref{fig2} (b) illustrates this relationship between $V_{S}$ and $V_{GS}$ for this branch. For the second branch, $V_{dd}$-M1-SPAD anode, $V_{S}$ can be solved from the current balance equation, which is: $i_{1} + \frac{V_{dd} - V_{S}}{R_{M1}} = \frac{V_{S}}{R},$ where $R_{M1}$ is the resistance of M1, which is determined by $V_{GS}$, and $R$ is the resistance between the SPAD anode and GND. When $V_{GS}$ is low, $R_{M1}$ is large, so $V_{S}$ is nearly 0 V. As $V_{GS}$ approaches the threshold voltage ($V_{on}$) of M1, $R_{M1}$ drops to approximately 1 $\Omega$, and $V_{S}$ approaches $V_{dd}$. This trend is shown by the blue line in Fig.~\ref{fig2} (b). In the stable idle state, the circuit reaches equilibrium at the intersection of the red and blue curves, where $V_{GS}$ is just below $V_{on}$ and $V_{S}$ is slightly less than $V_{dd}$. In the circuits, the $V_{Zener}$ is 4.3 V and the $V_{on}$ is $\sim$2.5 V.

When the QUENCHING signal transitions from HIGH to LOW, $V_{S}$ rises from 0 V to a high voltage, switching the SPAD from the armed to the idle state. The rise time of $V_{S}$ is a critical factor that limits both the maximum count rate and the gate-off speed. To accelerate this transition, an inductor L1 is employed. In the armed state, M2 conducts, allowing current to flow from $V_{dd}$ through R1, L1, and M2 to GND. When M2 turns off, L1 momentarily maintains the current via freewheeling through D1, causing $V_{GS}$ to briefly reach $V_{Zener}$. This turns M1 fully on, significantly boosting the rise speed of $V_{S}$.

Conversely, during the transition from the idle to the armed state, the QUENCHING signal switches from LOW to HIGH, rapidly turning on M2 and switching off M1. However, due to the high impedance between the SPAD anode and GND in this state, the anode voltage decreases slowly. To accelerate recovery, the RESET signal is briefly asserted, activating M3, which quickly pulls the anode voltage to GND. This active reset also occurs during the gate-on process.

\begin{figure*}[htbp]
\centerline{\includegraphics[width=15 cm]{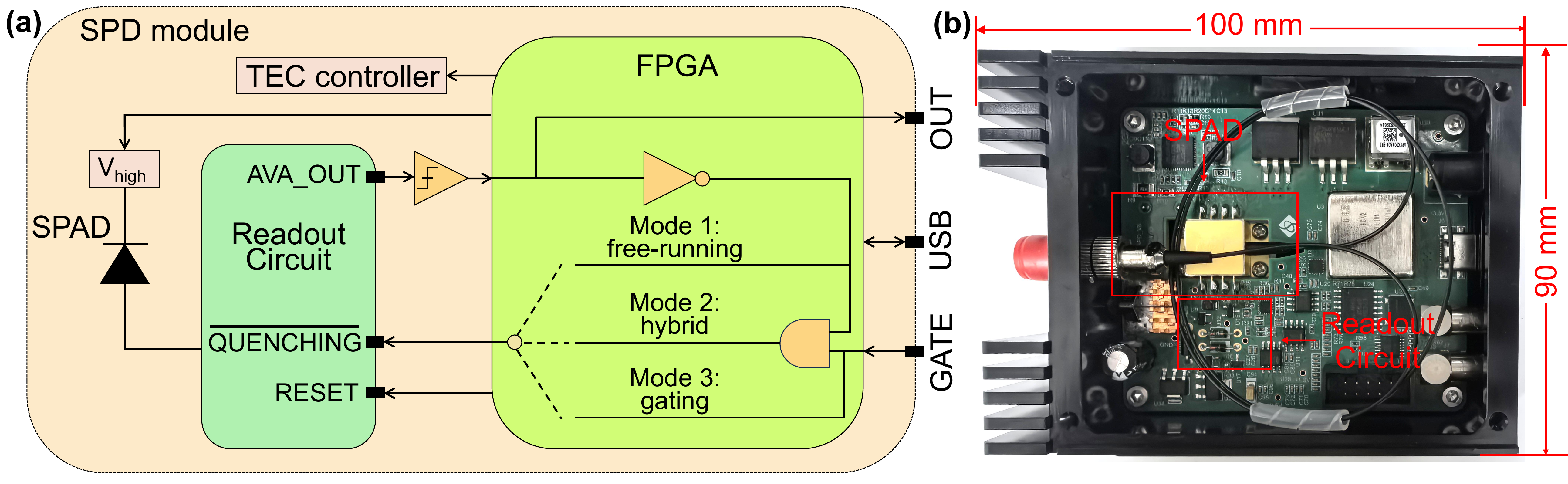}}
\caption{(a) The design diagram of the Si SPD module. (b) The photograph of the SPD module.}
\label{fig3}
\end{figure*}

\begin{figure*}[b]
\centerline{\includegraphics[width=18 cm]{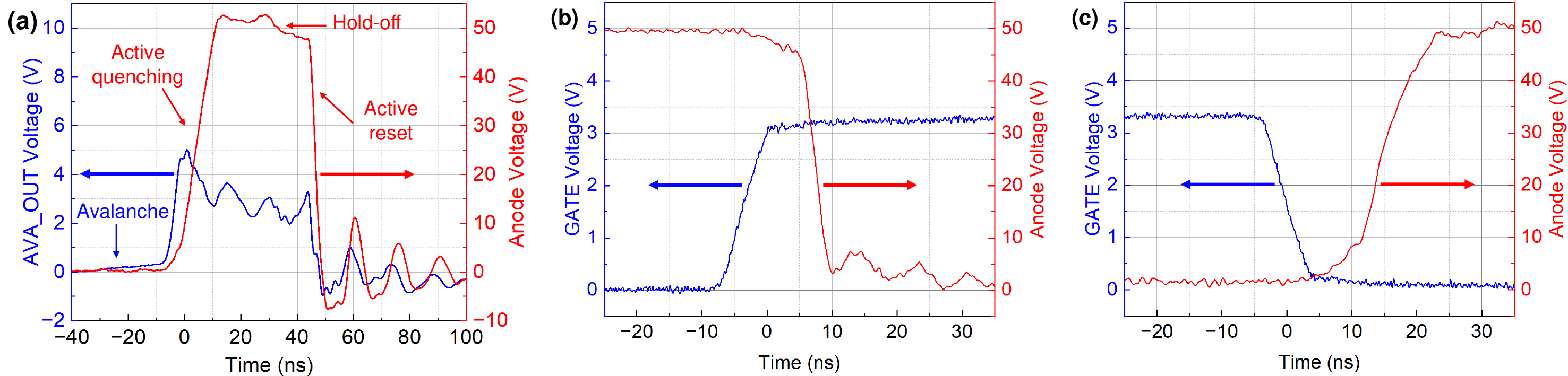}}
\caption{(a) The typical waveform of SPAD anode and AVA\_OUT during active quenching and active reset procedure in free-running mode. (b) The typical waveform of SPAD anode during the transition from gate-off to gate-on state. (c) The typical waveform of SPAD anode during the transition from gate-on to gate-off state.}
\label{fig4}
\end{figure*}

The avalanche extraction circuit comprises two DMOS transistors (M4 and M5) and a 1 k$\Omega$ sampling resistor (R3). The gate of M4 is biased at a fixed voltage, $V_{cc}$. In the armed state, both the SPAD anode and AVA\_OUT are at low voltage, keeping M4 fully conductive. When an avalanche event occurs, current flows through M4 and R3, resulting in a voltage rise at AVA\_OUT. This voltage change enables the detection of avalanche events by a downstream discriminator circuit. In the idle state, the SPAD anode voltage elevated, pushing the device out of Geiger mode. However, when the AVA\_OUT voltage approaches $V_{cc}$, M4 turns off, preventing further voltage rise. This ensures that the AVA\_OUT voltage remains below $V_{cc}$, thereby protecting the low-voltage discriminator from potential overvoltage damage. Similar to M3, M5 accelerates recovery by actively discharging the circuit after an avalanche event. 

\section{SPD Module}

Based on the Si SPAD and the readout circuit, we design a high-efficiency SPD module with multiple operation modes. Fig.~\ref{fig3}(a) presents the system diagram. A digitally adjustable high voltage ($V_{high}$) is applied to the SPAD cathode, while the anode interfaces with the readout circuit. In our system, the readout circuit operates with $V_{dd}$ = 53 V and $V_{cc}$ = 5 V. The output avalanche signal is translated to a low-voltage transistor-transistor logic standard. An FPGA governs the system, managing functions such as $V_{high}$ adjustment, TEC power control, and communication with a personal computer (PC) via a Universal Serial Bus (USB). The SPAD, readout circuit, and affliated circuits are integrated into a compact SPD module measuring 9 cm $\times$ 10 cm $\times$ 3 cm, weighting 380 g, and dissipating $\sim$ 8.5 W. A photo of the module is shown in Fig.~\ref{fig3} (b).

The SPD's operation mode is controlled by the FPGA and can be configured by the user through PC software. In free-running mode, the QUENCHING and RESET signals are triggered by the inverted detection signal. Upon receiving an avalanche signal, the FPGA immediately pulls the QUENCHING signal to GND, placing the detector in an idle state to promptly quench the avalanche. After a hold-off time of $\sim$50 ns, the QUENCHING signal is set HIGH, and the RESET signal is activated for about 10 ns to restore the SPD. In gating mode, the QUENCHING signal is directly driven by an external GATE signal, while the RESET signal is asserted for $\sim$10 ns following the rising edge of the GATE. In hybrid mode, a logical AND operation is performed between the inverted detection signal and the external GATE signal, and the result drives the QUENCHING signal. This allows the SPD to detect signal photons in free-running mode during the gate-on period, while exiting Geiger mode physically during the gate-off period.

The SPAD anode voltage is then measured in both free-running and gating modes. In free-running mode, the AVA\_OUT and SPAD anode waveforms are shown in Fig.\ref{fig4} (a). When an avalanche occurs, current flow causes a gradual voltage rise at AVA\_OUT. After approximately 20 ns, the avalanche is detected and active quenching is triggered, pulling the SPAD anode voltage up to 50 V. Due to the protective function of M4, the AVA\_OUT voltage is limited to a maximum of 5 V. Following a hold-off period of $\sim$50 ns, the SPAD anode is reset to 0 V, and the SPD is ready for the next photon detection. In gating mode, Fig.\ref{fig4} (b) illustrates that when the gate transitions from off to on, the anode voltage drops from 50 V to 0 V, with a delay of around 10 ns and a fall time of $\sim$5 ns. Conversely, Fig.\ref{fig4} (c) shows that during the gate-off transition, the anode voltage rises from 0 V to 50 V, with a delay of $\sim$15 ns and a rise time of about 10 ns, effectively forcing the SPAD to exit Geiger mode physically. 


Following standard calibration~\cite{JMH15}, the module is characterized in free-running mode under various operation temperatures. A picosecond pulsed laser diode at 785 nm with 100 kHz repetition rate serves as the photon source, emitting pulses with a full width at half maximum (FWHM) of less than 70 ps. To attenuate the laser intensity to an average of one photon per pulse, a variable optical attenuator, a 99:1 beam splitter (BS), and a fixed optical attenuator are employed. The 99\% port of the BS is monitored by a power meter to track the laser output. Detection signals from the module are captured by a time-to-digital converter (TDC) with 10 ps resolution, and the data is transferred to a personal computer to calculate parameters including PDE, DCR, afterpulse probability, and timing jitter. 

Assuming Poisson statistics for the laser pulses, the PDE is calculated as $PDE=-\frac{1}{\mu}\ln(1-\frac{R_{ph}}{f}),$ where $\mu$ is the mean photon number per pulse, $R_{ph}$ is the photon detection rate, and $f$ is the laser repetition frequency. The DCR is measured as the SPD's count rate in complete dark environment. The afterpulse probability is computed as $P_{ap}=R_{ap}/R_{ph}$, where $R_{ap}$ represents the afterpulse events derived from the TDC data. 

\begin{figure}[htbp]
\centerline{\includegraphics[width=8 cm]{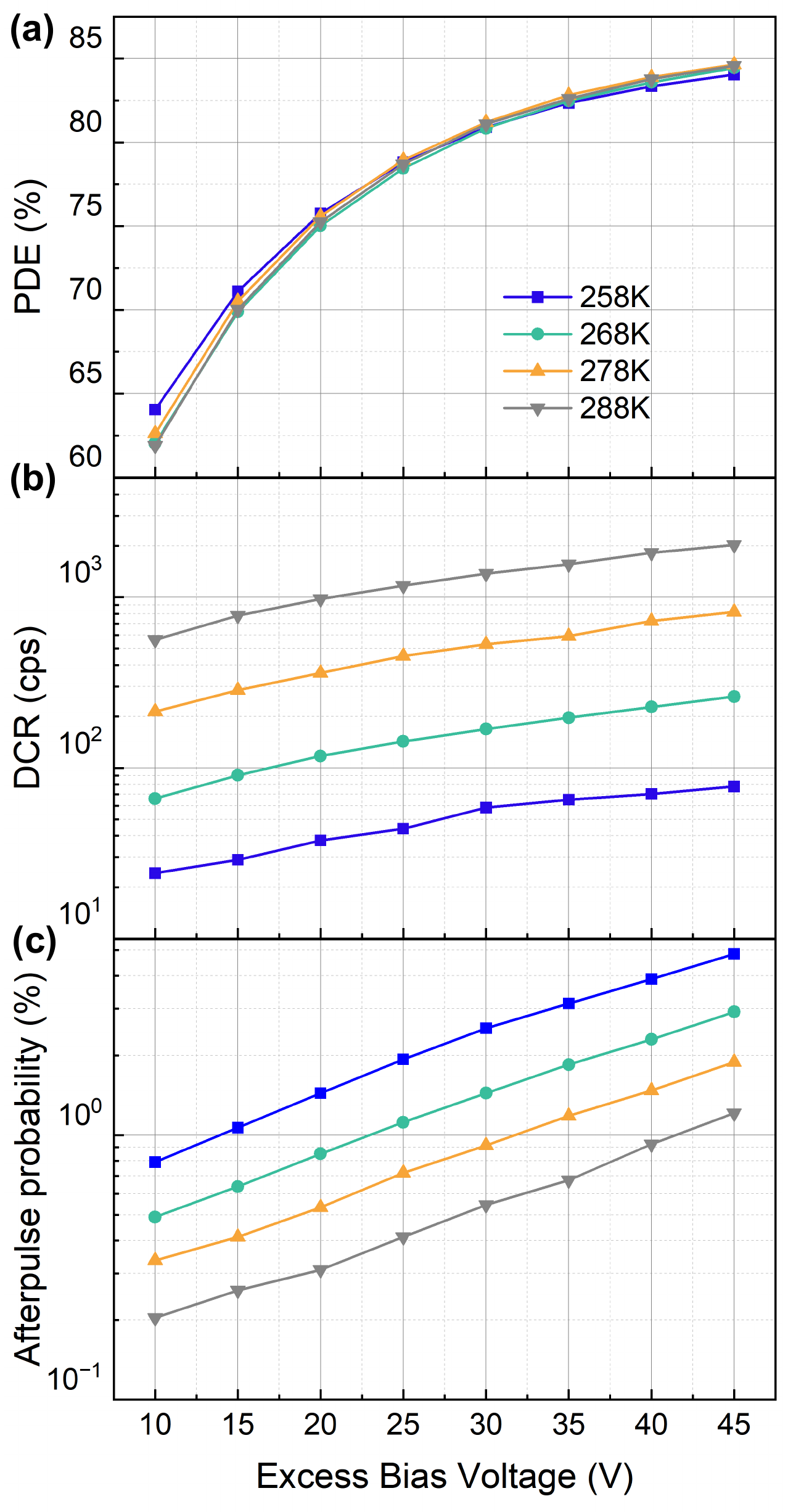}}
\caption{Measured (a) PDE, (b) DCR and (c) afterpulse probability versus excess bias voltage at different operation temperature.}
\label{fig5}
\end{figure}

The measured results for PDE, DCR and afterpulse probability versus excess bias voltage are presented in Fig.~\ref{fig5}. Benefiting from the optimized Si SPAD design and the high-amplitude quenching circuit, the Si SPD achieves a remarkable maximum PDE of up to 84\% at 785 nm under a 45 V excess bias voltage. At this point, the avalanche probability is nearly saturated, and further increasing the excess bias voltage does not lead to significant PDE improvement. The DCR and afterpulse probability exponential increase with excess bias voltage. At an operation temperature of 268 K, the DCR and afterpulse probability are 260 cps and 2.9\% with a PDE of 84.4\%. Cooling the SPAD to 258 K reduces the DCR to 80 cps, while increasing the temperature to 288 K lowers the afterpulse probability to 1.2\%. Compared to commercial products~\cite{SPCM}, the overall performance of the proposed Si SPD has shown significant improvement.

\begin{figure}[htbp]
\centerline{\includegraphics[width=8 cm]{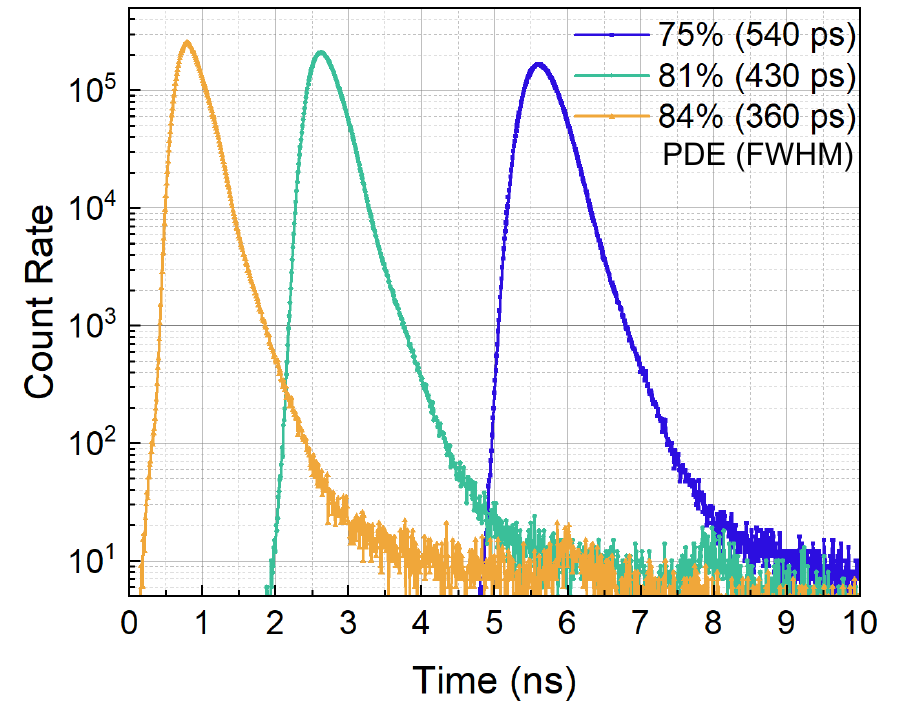}}
\caption{Total timing jitter measurements of the Si SPD at 268 K, measured with a laser width of $\sim$70 ps and a characterization system jitter of $\sim$30 ps.}
\label{fig6}
\end{figure}

Timing jitter is another critical performance metric for Si SPDs. Fig.~\ref{fig6} shows the measured timing jitter at various PDE levels at an operation temperature of 268 K. With a laser width of $\sim$70 ps and a characterization system jitter of $\sim$30 ps, the total full width at half maximum (FWHM) timing jitter of the Si SPD are 540 ps, 430 ps, and 360 ps at PDEs of 75\%, 81\%, and 84\%, respectively. The photon counts exhibit an asymmetric time distribution, which can be attributed to photons being absorbed in the neutral region and subsequently diffusing into the active region, creating a diffusion tail~\cite{MAI07}. Meanwhile, due to the faster rise of the avalanche signal at high excess bias voltage, the signal reaches the threshold more quickly, resulting in a propagation delay reduction of $\sim$5 ns as the PDE increases from 75\% to 84\%.

\section{Conclution}

In conclusion, we have developed a high-efficiency Si SPD module with multiple operation modes. The Si SPAD structure and epitaxial fabrication process are carefully optimized to maximize PDE and reduce noises. A novel readout circuit with a 50 V quenching voltage is developed to push the PDE of the SPD to its limit and enable flexible operation in free-running, gating, and hybrid modes. After calibration, the SPD achieves a PDE of 84.4\%, a DCR of 260 cps, an afterpulse probability of 2.9\%, and a timing jitter of 360 ps at 268 K. This work offers a practical solution for applications demanding ultra-high-efficiency Si SPDs with flexible operation modes.

\section*{Acknowledgment}

This work is supported by the Innovation Program for Quantum Science and Technology (2021ZD0300800, 2021ZD0300804), the National Natural Science Foundation of China (62175227, 62405305) and the Fundamental Research Funds for the Central Universities (WK9990000161).

\ifCLASSOPTIONcaptionsoff
\newpage
\fi






\end{document}